\input psfig.tex 
\documentstyle[11pt,twoside,headerfo]{article}
\renewcommand{\author}{Martin Pohl} 
\newcommand{\stitle}{Diffuse $\gamma$-rays from galactic halos}%
\renewcommand{\title}{Diffuse $\gamma$-rays from galactic halos}  
\def\gr{$\gamma$-ray }
\def\grs{$\gamma$-rays }
\begin{document}                                 
\setcounter{page}{1}                      
\pagestyle{empty}
\pageheaderlinetrue
\oddpageheader{}{\stitle}{\thepage}
\evenpageheader{\thepage}{\author}{}
\thispagestyle{empty}
\noindent
\vskip1cm
\begin{center}
\LARGE{\bf \title} 
\\[0.7cm]
\large{\author}         
\end{center}
\vspace{0.3cm}
\setcounter{section}{0}
\section{Introduction}
The \gr sky as we know it today is a composite of the emission
from point sources and diffuse emission (see Fig.1).
The most prominent feature is the galactic plane, in which interactions
between cosmic rays and the thermal gas lead to \gr emission by $\pi^0$-decay
and bremsstrahlung. On top of this diffuse emission we see point sources
all over the map, part of them being pulsars, another part distant
AGN, and also a significant fraction of yet unidentified sources. But we
also see that the diffuse emission extends out of the galactic plane
up to the poles. There appears to be an isotropic emission component which is 
presumably extragalactic in origin and may be understood as blend of
unresolved AGN, but there is also considerable galactic emission.
At higher latitudes interactions between cosmic rays and
thermal gas still play a role, but inverse-Compton scattering of ambient
photons by cosmic ray electrons becomes increasingly important.
This emission tells us about the physical conditions in the halo
as seen by the cosmic ray particles, and it may also reveal
previously hidden gas, e.g. baryonic dark matter.

\begin{figure}[htb]
\psfig{figure=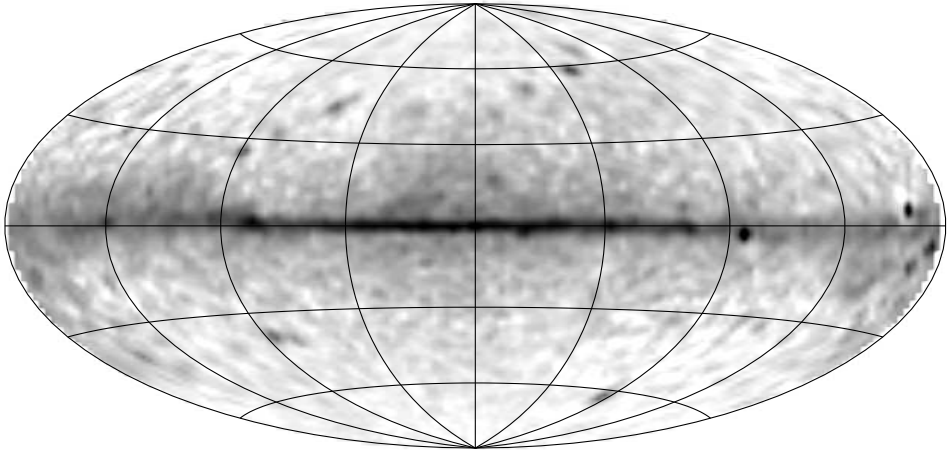,width=13.0truecm,clip=}
\caption{The EGRET sky above 100 MeV \gr energy. The image is deconvolved 
by a Maximum-Entropy algorithm (for the method see Strong 1995).
The grey scale is logarithmic
with darkness indicating high intensity.}
\end{figure}

\section{The r\^ ole of confused point sources}
Locally, any analysis of the galactic diffuse emission can be seriously
hampered by unresolved galactic point sources which may have a sky distribution
similar to that of gas. As is shown in Table 1, in the galactic plane the
source density is significantly higher than at high latitudes, not only in
total number but especially for the unidentified sources.
If the sky distribution
of sources were isotropic we would expect a smaller source density
in the plane, since the strong background there reduces 
the statistical prominence of any source even for the on average
higher exposure. A modelling of the
sky distribution of sources under the assumption that all
unidentified sources are galactic has revealed that unresolved 
galactic sources would account for 30-40\% of the total galactic
\gr luminosity above 100 MeV and that the sky distribution
of the unresolved sources could resemble that of the gas since we
see mainly the more distant sources (Kanbach et al. 1996). Although
in reality the numbers may not be that bad, perhaps around 20\%, for
halo studies the point source contribution has to be taken into account.
 
\begin{table}[htb]
\begin{tabular}
{|r|c|c|c|c|c|}\hline
$|$b$|$ & [sr] & Identified & Unidentified &Identified/Total & Total/steradian\\ 
& & & & & \\ \hline
0-10 & 2.2 & 5 & 32 & 0.14$\pm\,$0.06 & 16$\pm\,$3.0\\
10-30 & 4.1 & 18 & 26 & 0.41$\pm\,$0.11 & 11$\pm\,$1.6\\
30-90 & 6.3 & 32 & 14 & 0.70$\pm\,$0.16 & 7.3$\pm\,$1.1\\
\hline
\end{tabular}
\caption{Source statistic on the basis of the second
EGRET catalog (Thompson et al. 1995).
In the galactic plane the source density is higher than at high latitudes
and also the fraction of identified sources is unusually small.}
\end{table}

If there were no point sources the \gr intensity in the
galactic plane should follow locally the gas distribution.
In fact the
EGRET team uses a scaling model on the basis of the gas distribution
as null hypothesis in the search for point sources (Hunter et al.
1996). We can test the reliability of such models on small scales
by comparing it to deconvolved data for the galactic center region
where the statistic is best, i.e. small scale structures have the 
highest statistical significance. We will
concentrate on data at higher energies where EGRET's point
spread function is smaller and less photons are required to
reproduce the intensity structure. The result is shown in Fig.2
and Fig.3, where despite the larger pixel size in the model (Fig.3)
it is obvious that there is a wealth of structure not
directly related to the gas. Part of this additional structure
is listed as unidentified sources in the second EGRET catalog.

It should be noted that comparing deconvolved data to the original
data is more sensitive to small scale discrepancies than the
usual comparison of a convolved model to original data, especially 
if the global sky distribution of point sources is similar to the distribution of gas. Therefore, our finding does not contradict
the statement of Hunter et al. (1996) that the gas coupling model, 
convolved and averaged over strips a few degrees wide, is a good
representation of the observed sky distribution at \gr energies
above 100 MeV.

\section{Diffuse emission in the EGRET range}
\subsection{The \gr emissivity}
In galactic halos inverse-Compton scattering (IC) of ambient photons by
cosmic ray electrons is an important production process for \grs .
Due to the large scale heights of far-infrared photons and the cosmic
microwave background, the latitude distribution of this
emission is much broader than in case of the bremsstrahlung
and $\pi^0$-decay. Thus the IC component can tell us
about the lifetime of cosmic ray electrons in the Galactic halo.
\begin{figure}[ht]
\psfig{figure=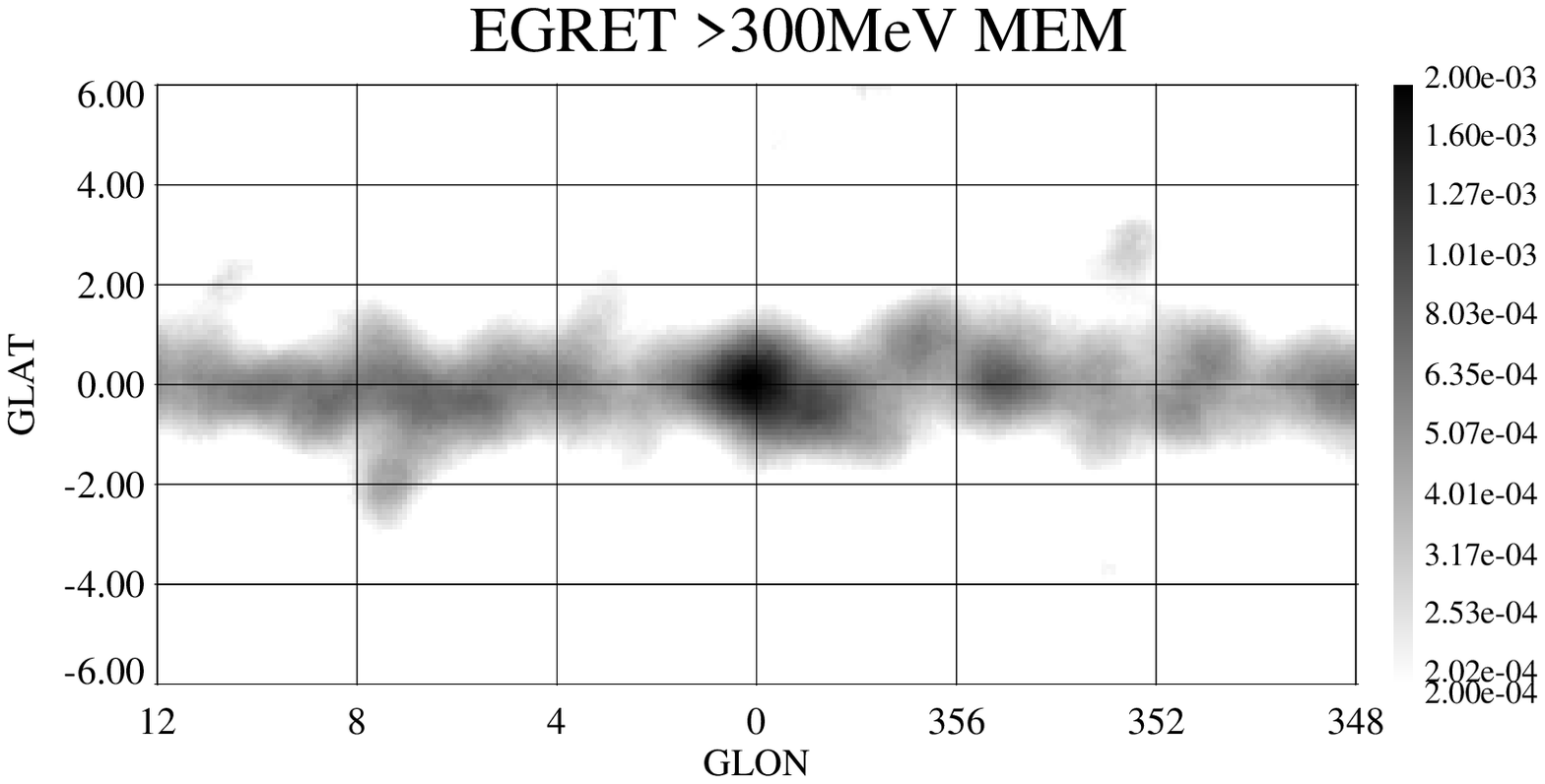,width=13.1truecm,clip=}
\caption{The galactic center region above 300 MeV \gr energy. The 
image is deconvolved by a Maximum-Entropy algorithm. The
intensity scale is in units ${\rm cm^{-2}\, sec^{-1}\, sr^{-1}}$.}
\psfig{figure=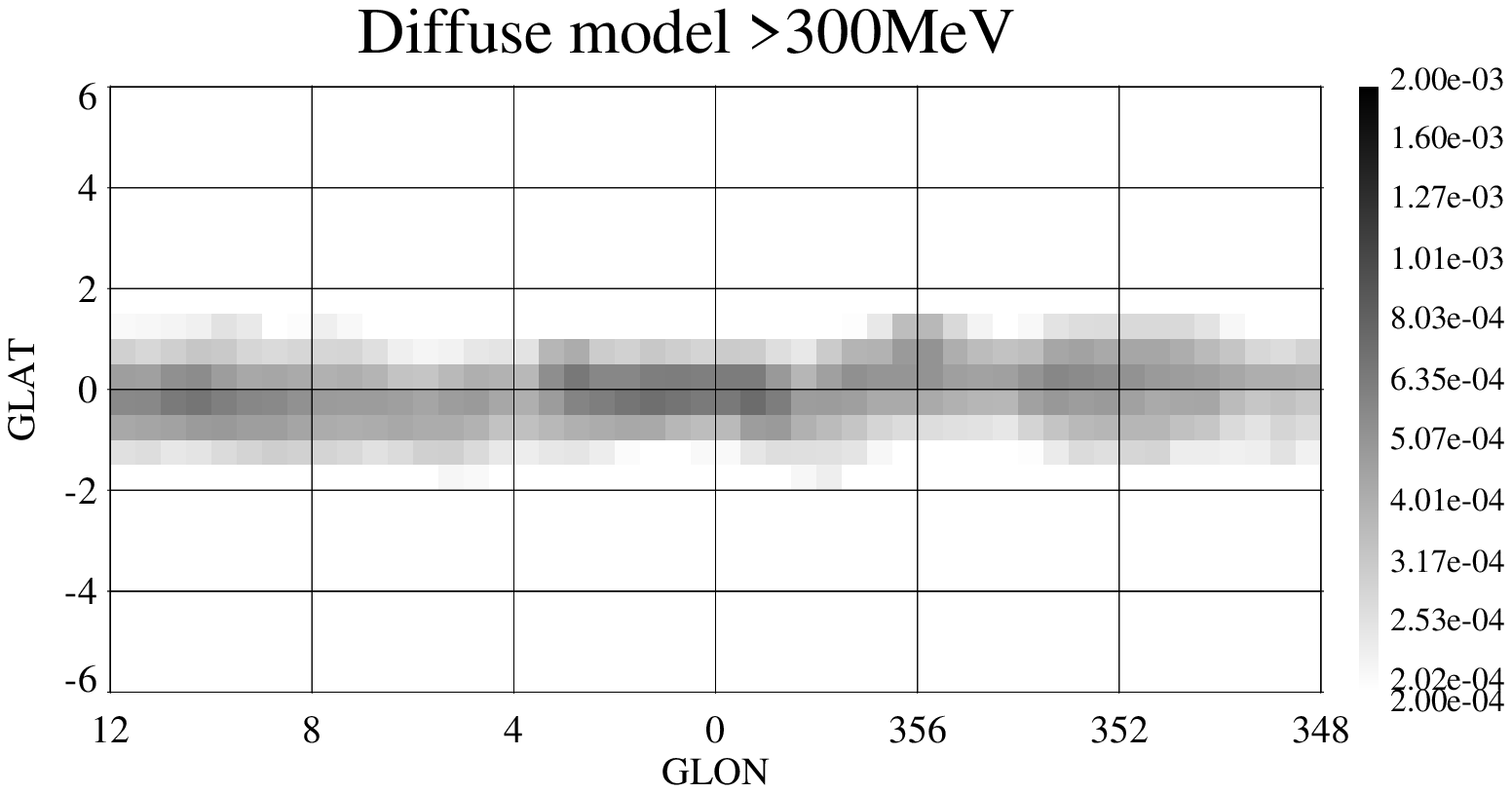,width=13.1truecm,clip=}
\caption{The model prediction based on gas distribution for the 
galactic center region. The scale is the same as for the deconvolved EGRET data in Fig.2. The disagreement between Fig.2 and Fig.3 is obvious, implying that point sources contribute significantly.}
\end{figure}
Unfortunately the IC emission is generally weak and has a spectral shape
similar to that of the extragalactic \gr background. However, the latter
should be isotropic, and one may use the contrast between
inner and outer Galaxy to detect the IC component.

One may also compare the spatial distribution of \grs not
related to gas to the brightness temperature distribution in radio surveys. Depending on frequency the bulk of radio emission at
high latitudes will be due to synchrotron radiation of cosmic ray
electrons, and one should expect correlations between the synchrotron
flux density and the IC intensity, although the typical
electron energy for synchrotron emission at 408 MHz is roughly an 
order of magnitude less than the energy required for up-scattering
infrared photons. Such work has been done by Chen et al. (1996),
who conclude that the average intensity of IC
emission at high latitudes is $(5\pm 0.8)\cdot 10^{-6}\ {\rm 
cm^{-2}\, sec^{-1}\, sr^{-1}}$ for $E>100\,$MeV (around
one third of the diffuse extragalactic background) and that the 
photon spectral index is $-1.85\pm 0.17$.
\begin{figure}[htb]
\psfig{figure=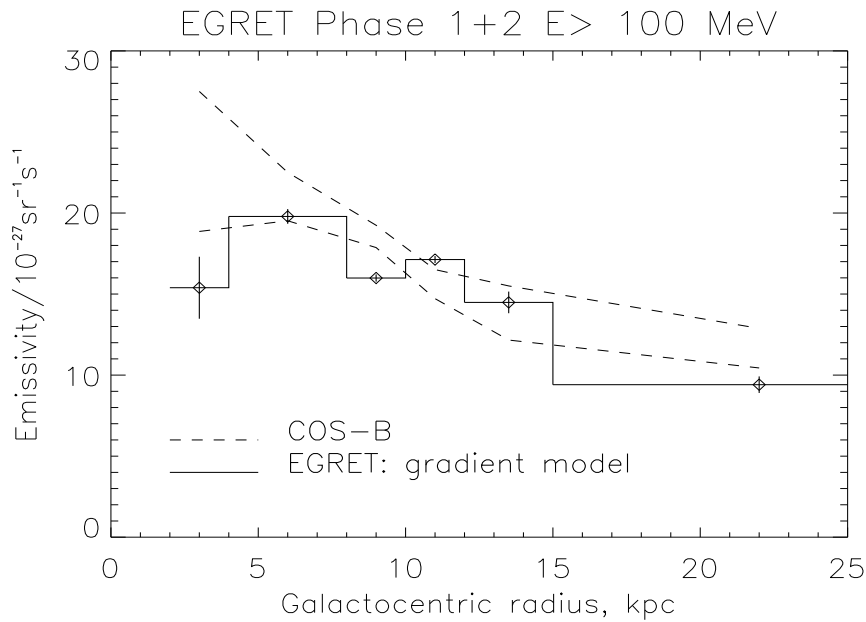,width=13.1truecm,height=8.5truecm,clip=}
\caption{The \gr emissivity per H-atom averaged over galactocentric
rings. There is weak decrease of emissivity with galactocentric
radius, the so-called gradient, basically confirming the earlier COS-B result.
(Strong and Mattox 1996)}
\end{figure}

To investigate the \gr emission originating from $\pi^0$-decay and 
bremsstrahlung we need some prior knowledge on the distribution of gas
in the Galaxy. This includes not only H$I$ but also H$_2$, which is 
indirectly traced by CO emission lines, and H$II$, which is traced by
H$\alpha$ and pulsar dispersion measurements. Even in case of the directly 
observable atomic hydrogen we obtain only line-of-sight integrals,
albeit with some kinematical information. Any deconvolution of the velocity
shifts into distance is hampered by the line broadening of individual
gas clouds and by the proper motion of clouds with respect to the
main rotation flow. The distance uncertainty will in general be around
1 kpc. 

It may be appropriate to use only a few resolution
elements and investigate the \gr emissivity per H-atom in galactocentric
rings, i.e. to assume azimuthal symmetry for the emissivity. This kind
of approach has already been successfully applied to the earlier COS-B data.
Strong and Mattox (1996) have repeated the analysis on EGRET data
at \gr energies above 100 MeV. These authors report that
there is a gradient, that is a decline of the \gr emissivity per
H-atom with galactocentric radius, confirming earlier COS-B results.
This radial gradient is weak and does not exceed a factor 2 difference
between inner and outer Galaxy (see also Fig.4). It should be noted
that this gradient does not necessarily hold locally. A comparison of
the \gr emissivity in the perseus arm at 3 kpc distance in the outer galaxy to
the Cepheus and Polaris flare at 250 pc gave a difference of a 
factor 1.7$\pm$0.2, by far exceeding the overall gradient (Digel et al.
1996).

\subsection{Spectral information}
The \gr production processes may not only be separable
by their different sky distribution, but also by their spectral shapes.
While the $\pi^0$-decay should result in a broad line centered on 68 MeV,
the bremsstrahlung spectrum should follow the rather
steep electron spectrum, thereby partly hiding the signature of the
$\pi^0$-decay line. The emission due to IC scattering is
expected to have a hard spectrum, similar to that of the diffuse
extragalactic background. Strong and Mattox (1996) have performed
a spectral analysis of EGRET data in the ten standard energy bands
to investigate the spectrum of the isotropic extragalactic emission, the
emission correlated to the gas distribution ($\pi^0$-decay and
bremsstrahlung), and the emission due to IC scattering,
for which the expected sky distribution was modelled.
The result is shown in Fig.5 together with the overall spectrum of diffuse
emission in EGRET data compared to the results of COMPTEL and the earlier
mission COS-B. All sources in the second EGRET catalog have been taken
into account in this analysis.
\begin{figure}[htb]
\psfig{figure=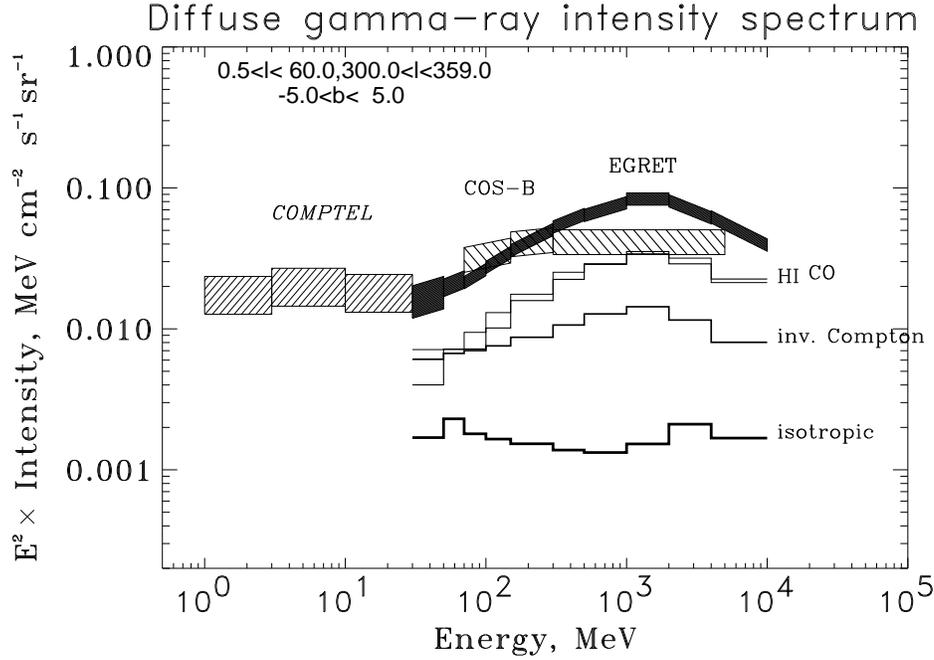,width=13.1truecm,clip=}
\caption{The \gr intensity in EF(E) representation. Shown is both the total spectrum (black boxes) and the spectra of the individual components
(as histograms) for the inner Galaxy.
For comparison the COMPTEL results at lower energies and the old COS-B
data are included. The uncertainties in the spectrum of the individual 
components are around 10\% except for the low-energy points.
(Strong and Mattox 1996)}
\end{figure}

The spectrum of the isotropic extragalactic component is around $E^{-2}$, 
harmonizing with the average AGN spectrum. The IC emission has a
slightly harder spectrum similar to the results of Chen et al. (1996) at 
higher latitudes. However, the spectrum of emission related to the gas is 
difficult
to understand. It deviates significantly from the expected superposition
of $\pi^0$-decay and bremsstrahlung. 
At energies above 1 GeV there is a clear excess of emission related to gas
which may be somewhat relaxed by variation of the input proton spectrum
to the pion production process. Part of this excess may also be explained
by unresolved sources, however the small intensity at low energies requires
sources with very hard spectrum and possibly a cut-off at a few GeV. Only old pulsars like Geminga have a corresponding spectrum and may also have a sky distribution resembling that of gas.
\begin{figure}[htb]
\psfig{figure=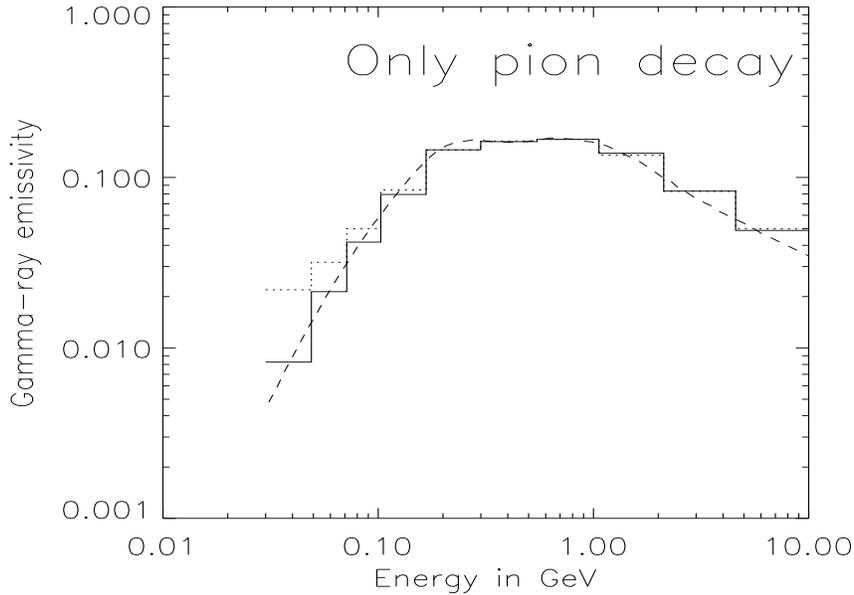,width=13.1truecm,height=8.0truecm,clip=}
\caption{A demonstration of the spill-over effect at low energies.
The dashed curve is the true input spectrum in EF(E) representation
of arbitrary units, here a $\pi^0$-decay spectrum.
The solid line histogram is the corresponding average over the standard
EGRET energy bands. The dotted histogram shows what the result of
the standard EGRET analysis would be. At low energies the spectrum
is overestimated by a factor of three.}
\end{figure}

The spectral hardness of the emission related to gas at low energies is
also not easy to understand. The standard analysis tools of EGRET
for diffuse emission implicitly
assume a power-law spectrum of photon index $2.1$. If the true intensity
is softer than this the standard analysis will underestimate the emission
at low energies, while for hard spectra the emission will be overestimated.
The reason for this is that a significant part of the low energy counts are due to misinterpreted photons of higher true energy. An example of this
effect is shown in Fig.6. It turns out that even in the absence of any
sky emission below 100 MeV the spill-over provides roughly an $E^{-1}$ 
differential photon spectrum at these energies. The spectrum of the 
emission related to gas in Fig.5 is not far from this, indicating that
the true sky intensity at low energies is even less. This would imply
that the density of 100 MeV electrons is relatively small compared to the
density of 10 GeV electrons which are responsible for the IC
emission, considering also the radio synchrotron data which tell about
the electron spectrum in the GeV range.

\section{Diffuse emission at hard X-rays and soft \grs}

At lower energies we have data from three instruments for the inner Galaxy: 
COMPTEL in the MeV range (Strong et al. 1994), OSSE in the 100 keV range 
(Purcell et al. 1996), and GINGA around 10 keV (Yamasaki et al. 1996). The 
resulting spectrum of diffuse emission from 2 keV to 30 MeV is shown
in Fig.7. 

The GINGA spectrum may partly be thermal in origin. In fact 
a thermal model plus absorbed power-law fits well to the data. 
Also note the prominent Fe K$\alpha$ line at 6.7 keV.
While the point source contribution in the COMPTEL range is probably 
around 20\% as in case of EGRET, the confusion in the OSSE range is
less easy to estimate. OSSE is a non-imaging instrument and therefore
cannot identify point sources on its own. The galactic center region has
been simultaneously observed
by SIGMA, a coded-mask instrument designed to search for sources. It
turned out that after subtracting the flux due to point sources
resolved by SIGMA the residual OSSE diffuse emission was similar to
the total OSSE flux at l=25$^\circ$, which supports the view that 
the residual is extended diffuse emission (Purcell et al. 1996).
We show only the residual flux in Fig.7, for which
still the contamination by unresolved sources is hard to estimate,
but is unlikely to account for all the emission.
\begin{figure}[htb]
\psfig{figure=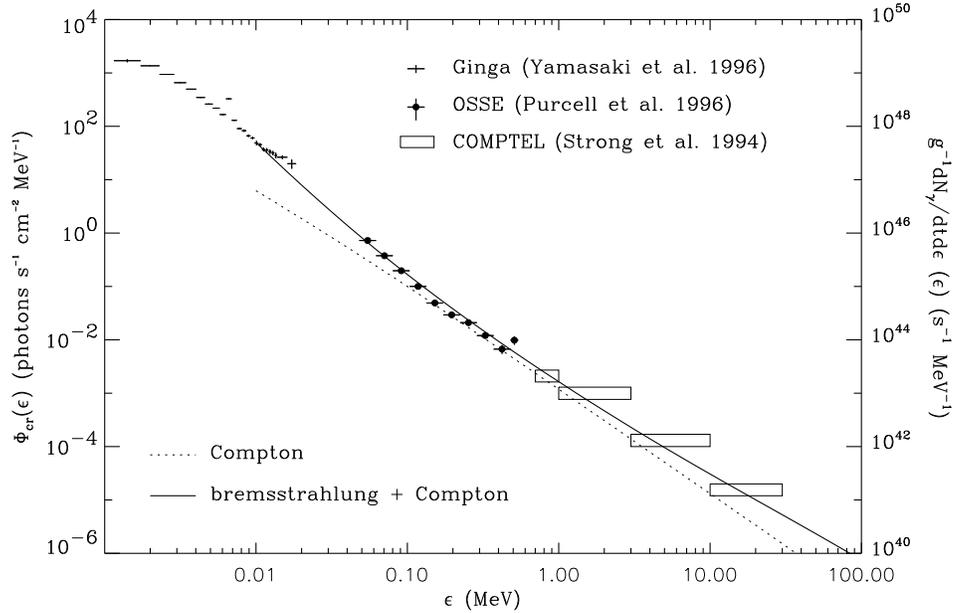,width=13.1truecm,clip=}
\caption{The \gr intensity in direction of the inner Galaxy
from X-rays to soft \grs. The GINGA spectrum may
partly be thermal in origin. However, the OSSE data require a very soft
electron spectrum at low energies. The solid line is a model curve of 
bremsstrahlung and IC scattering based on such a soft electron 
spectrum. (Skibo et al. 1996)}
\end{figure}

The energy dependence of Coulomb and ionisation losses of electrons in
the MeV range imply that they cannot travel far from their sources and
that their spectrum should be hard. Thus the OSSE result can only be
explained by invoking an additional source of low energy electrons with an
input power exceeding that provided by Galactic supernovae. Skibo et al.
(1996) have argued that this phenomenon may be transient due to Galactic
spiral density waves, thereby relaxing the energy requirement. 
Very promising appear models which explain the soft spectrum of low
energy electrons as direct result of stochastic acceleration out of the
thermal pool working against Coulomb interactions
(Schlickeiser, this volume). The energy input in this class of models
would come from interstellar turbulence.

The existence of a high density of low-energy electrons is an important
issue for the energy and ionisation balance of the interstellar medium
both in the plane and in the halo,
since the bulk of the energy input is directly transferred to the thermal gas.

\section{The Magellanic Clouds}

The Magellanic Clouds are two irregular galaxies located at around 50 kpc
distance. The Large Magellanic Cloud (LMC) has been detected by EGRET
(see also Fig.8), while for the Small Magellanic Cloud (SMC) EGRET has
derived only an upper limit (Sreekumar et al. 1993).
Both in LMC and in SMC the \gr emissivity per H-atom is considerably 
smaller than in our Galaxy. This result implies that the bulk of
GeV cosmic rays is galactic in origin. Since
the gas in LMC and SMC is much less illuminated by cosmic rays than gas
in the Galaxy, the cosmic ray density at some distance from the Galaxy
may be very small. This has impact on dark matter studies, since this
fact allows a substantial amount of baryonic dark matter to be hidden
at a few 10 kpc from the Galaxy without inducing observable \gr emission.
\begin{figure}[htb]
\psfig{figure=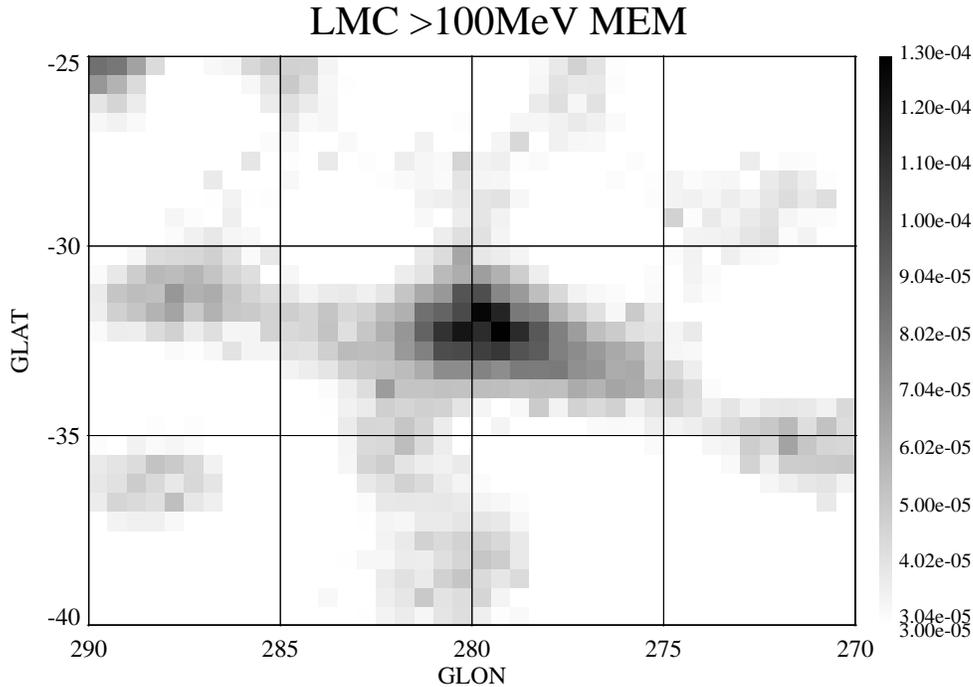,width=13.1truecm,clip=}
\caption{The \gr intensity above 100 MeV for the region of LMC. The data
are deconvolved again and the scale is in ${\rm cm^{-2}\,sec^{-1}\,sr^{-1}}$.
LMC is clearly seen with its emission concentrated along the gas ridge south
of 30 Doradus.}
\end{figure}

There has been a debate on whether the \gr flux of LMC and SMC in relation
to their radio emission would allow equipartition between the magnetic field
and cosmic rays. It turns out that this equipartition is still possible 
provided one allows the cosmic ray electron-to-proton ratio to be different
from that in the solar vicinity. The \gr data of the Magellanic Clouds are
best explained when the e/p ratio is much higher so that basically only the
density of cosmic ray nucleons is reduced in LMC and SMC (Pohl 1993).

All results mentioned above were derived based on the integral properties
of the Magellanic Clouds. A new study including spatial and spectral
information is under way.

\section{What of the future?}
The next step in analysing \gr data of diffuse emission will be based on
propagation modelling of cosmic rays which better allows to link the 
\gr data to the results of radio observations and direct cosmic ray 
measurements.
Such models are underway and they may help to answer the question of
whether the sources of cosmic rays are localized entities, e.g.
supernova remnants, or whether they are diffuse, and in relation to this, 
whether reacceleration plays a r\^ ole.
With forthcoming instruments we also may be able to set more stringent
limits on the baryon fraction in a dark matter halo.

The ESA M2 mission INTEGRAL will not be ideally suited for studies of
diffuse \gr continuum emission from the halo as it is specifically designed
for high spatial and energy resolution. However, two project studies
currently undertaken are highly promising. The project GLAST will use 
silicon-strip detectors to measure \grs in the energy range 50 MeV to 100
GeV a factor 100 better in sensitivity than EGRET. At low energies of
1 MeV to 50 MeV laboratory studies are undertaken to combine the Compton
telescope principle with silicon-strip tracking of the scattered electron,
which may also give a factor 100 in sensitivity compared to
COMPTEL. So if we continue to invest in \gr astronomy we may go a big step 
forward in understanding diffuse \grs from Galactic halos in the next decade.

{\bf Acknowledgements:}
I am indebted to Andy Strong and Jeff Skibo who kindly provided 
figures and plots in electronic form, partly prior to publication.

\section*{References}

\begin{description}{}\itemsep=0pt\parsep=0pt\parskip=0pt\labelsep=0pt

\item Chen A., Dwyer J., Kaaret P.: 1996, ApJ, in press

\item Digel S.W., Grenier I.A., Heithausen A. et al.: 1996, ApJ, in press

\item Hunter S.D. et al.: 1996, A\&AS, in press

\item Kanbach G., Bertsch D.L., Dingus B.L. et al.: 1996, A\&AS,
in press

\item Pohl M.: 1993, A\&A 279, L17

\item Purcell W.R. et al.: 1996, A\&AS, in press

\item Skibo J.G., Ramaty R., Purcell W.R.: 1996, A\&AS, in press

\item Sreekumar P., Bertsch D.L., Dingus B.L. et al.: 1993, PRL 70, 127

\item Strong A.W.: 1995, Exper. Astron. 6, 97

\item Strong A.W. et al.: 1994, A\&A 292, 82

\item Strong A.W., Mattox J.R.: 1996, A\&A, in press

\item Thompson D.J., Bertsch D.L., Dingus B.L. et al.: 1995,
ApJS 101, 259

\item Yamasaki N. et al.: 1996, A\&AS, in press

\end{description}

\vskip1cm
\noindent

{\bf Address of the author:}\\[0.4cm]

{\sc Martin Pohl:}
Max-Planck-Institut f\"ur Extraterrestrische Physik, Postfach 1603,
85740 Garching, Germany; mkp@mpe-garching.mpg.de
\end{document}